\documentclass[aps, pra, 12pt, floatfix]{revtex4-2}

\usepackage{amsmath}
\usepackage{amssymb}
\usepackage{mathtools}
\usepackage{xcolor}
\usepackage{bm}
\usepackage{dcolumn}
\usepackage{graphicx}
\usepackage[export]{adjustbox}
\usepackage{booktabs}
\usepackage{multirow}
\usepackage{hyperref}

\newcommand\hl[1]{%
  \bgroup
  \hskip0pt\color{red!80!black}%
  #1%
  \egroup
}

\begin{document}

\title{Magnetic Jones Vector Detection with RF Atomic Magnetometers}

    \author{C.~Z.\ Motamedi}
    \email{cmotamed@gmu.edu}
    \affiliation{%
        Quantum Science and Engineering Center, George Mason University,
        Fairfax, VA 22030, USA%
    }
    \author{K.~L.\ Sauer}
    \email{ksauer1@gmu.edu}
    \affiliation{%
        Quantum Science and Engineering Center, George Mason University,
        Fairfax, VA 22030, USA%
    }

\begin{abstract}

    We show, theoretically and experimentally, how the absolute orientation and polarization state of radio-frequency (RF) magnetic fields in a transverse 2D plane can be uniquely determined using two optically pumped atomic magnetometers. In addition, the output signals from these quantum sensors can be readily expressed as a magnetic Jones vector. This composite device can complement electric field detection in finding RF directions, or it can be used in applications where the electric field is negligible. The latter is found in a myriad of applications where the source can be modeled as radiating magnetic dipoles in the near-field limit. This new tool could be used to characterize a material's response to RF excitation or to distinguish signal from noise.
    
\end{abstract}

\maketitle

\section{Introduction}

For distances $r$ away, a magnetic dipole radiating at frequency $\omega$ only produces a magnetic field in the near field limit ($r << 1/\omega\sqrt{\mu\epsilon}$)~\cite{ramo1994}. Many applications, such as nuclear magnetic resonance~\cite{savukov2007,lee2006,ledbetter2008,begus2017}, magnetic induction tomography~\cite{deans2021,fregosi2020,bevington2020,wickenbrock2014,marmugi2015,marmugi2016}, and imaging of material defects~\cite{bevington2019} naturally operate in this regime. Other applications, such as magneto-inductive communication~\cite{romanofsky2020,sojdehei2001}, purposely use low frequencies to create a large near-field to allow for penetration of the magnetic field into semi-conductive or conductive materials. Atomic magnetometers can have a critical role in such applications because of their improved sensitivity in the low-frequency regime compared to conventional coil detection~\cite{lee2006,keder2014,savukov2007,yao2022,cooper2022}, as well as their ability to independently map out magnetic fields~\cite{cooper2016,bevilacqua2019} without inductive coupling to each other or capacitive coupling to the environment. RF magnetometers have also been used for the detection of exotic fields through the GNOME collaboration~\cite{afach2018}. Many applications would benefit if the magnetometers could operate unshielded. However, ambient noise and interference can be prohibitively high. Some techniques that have been used to reduce noise in unshielded environments include active bias stabilization~\cite{deans2016, deans2018,yao2022} and an intrinsic gradiometer~\cite{cooper2022}.

For applications in communication, methods for reducing noise have been proposed that take advantage of the magnetometer’s sensitivity to the polarization state of the RF field. It has been demonstrated that atomic magnetometers have a $>$36 dB difference in sensitivity between oppositely rotating RF fields~\cite{gerginov2019}, and the sensitivity to fields that rotate in the same sense as the Larmor precession is twice that for linearly polarized fields~\cite{oida2012}. Because of this, a sensor that detects circularly polarized fields while rejecting linearly polarized noise has been proposed for communication applications~\cite{gerginov2019}. Such a sensor would be useful for a magnetic communication system~\cite{gerginov2017} that uses polarization modulation with circularly polarized signals. Polarization modulation could be used simultaneously with other types of modulation as a communication technique that would allow more information to be carried on the same channel~\cite{epstein1968,abidin2012}, and it has been suggested that the use of circularly polarized signals could reduce reflection and absorption at the interface of the electrically conductive medium for radio frequencies in the kilohertz range~\cite{fan2022}. Because of their sensitivity to the polarization state of the signal, atomic magnetometers can be used as the receiver for these signals~\cite{fan2022}. Besides applications in communication, the polarization state of the magnetic field can be used to determine orientation~\cite{maguire2015} or for radio direction finding.

The experiment presented in this paper takes advantage of the magnetometer’s sensitivity to the polarization state of the magnetic field for determining the absolute direction of a linearly or elliptically polarized RF field within a 2D plane. While DC magnetometers have been used to determine the absolute field direction~\cite{seltzer2004,patton2014}, similar efforts have not previously been made for RF fields. Such direction sensitivity can be used to separate out signal from noise. For instance, in nuclear quadrupole resonance (NQR) detection of buried explosives~\cite{garroway2001}, the signal of interest is orthogonal to the ground, but radio-interference is largely parallel to the ground~\cite{thomson1989} due to boundary conditions for far-field radiation. In this case, the direction could be used to distinguish between signal and interference for NQR detection of landmines.

In this work, we demonstrate how the polarization state of the magnetic field in a 2D plane can be uniquely determined and represented by a magnetic Jones vector~\cite{carozzi2009}. The technique presented here uses two crossed pump-probe magnetometers with colinear probe beams. Their static magnetic fields point in opposite directions, so that the two oppositely rotating circularly polarized components can be detected separately by the two magnetometers. Using phase-sensitive heterodyne detection, the polarization states of linearly, circularly, and elliptically polarized RF fields are determined from the equations
\begin{eqnarray}
    \widetilde{B}_x &\propto& S_1+S_2, \\
    \widetilde{B}_y &\propto& i(S_2-S_1),
\end{eqnarray}
where $\widetilde{B}_x$ and $\widetilde{B}_y$ are the $x$- and $y$-components of the RF field in phasor notation, and $S_1$ and $S_2$ are the complex signals from the two sensors.

\maketitle

\section{Theory}

The magnetization of the atoms as a function of time is given by the Bloch equations. The coordinate system in the lab frame is defined so that the $z$-axis points in the direction of the applied static field, or tuning field, $B_0$. In a coordinate system that rotates about the $z$-axis in the lab frame with angular frequency $\omega$,
\begin{eqnarray} \label{Bloch_comp}
\frac{\partial M_{x'}}{\partial t} &=& \frac{-M_{x'}}{T_2}-\left[\left(\gamma\textbf{B}+\omega\hat{z}\right)\times\textbf{M}\right]_{x'},\\
\frac{\partial M_{y'}}{\partial t} &=& \frac{-M_{y'}}{T_2}-\left[\left(\gamma\textbf{B}+\omega\hat{z}\right)\times\textbf{M}\right]_{y'},\\
\frac{\partial M_z}{\partial t} &=& \frac{-M_z}{T_1}-\left[\left(\gamma\textbf{B}+\omega\hat{z}\right)\times\textbf{M}\right]_z,
\end{eqnarray}
where $\gamma$ is the gyromagnetic ratio of the atoms, $\textbf{B}$ is the full magnetic field, and the relaxation time constants, $T_1$ for longitudinal and $T_2$ for transverse, are those after the termination of the pump beam. The rotating frame axes are denoted with primes, and the lab frame is left unprimed.
If, in addition, there is an RF field in the $xy$-plane, then $\textbf{B}$ can be written as
\begin{eqnarray}
\textbf{B}=f(\omega t)\hat{x}+g(\omega t)\hat{y}+B_0\hat{z},
\end{eqnarray}
where $f(\omega t)$ and $g(\omega t)$ are periodic functions with frequency $\omega$. Any component of the RF field parallel to the static bias field can be ignored in the weak field limit, when the strength of the RF field is much less than the bias field strength.

If we define $\omega_0\equiv -\gamma B_0$, ${h(\omega t)\equiv f(\omega t)+ig(\omega t)}$, and $M_\pm \equiv M_{x'}\pm iM_{y'}$, the magnetic field is given by
\begin{eqnarray}
2\textbf{B}=\left[h(\omega t) \left(\hat{x}'-i\hat{y}'\right)e^{-i\omega t}+B_0\hat{z}\right]+c.c.
\end{eqnarray}
in the rotating frame, where $c.c.$ stands for the complex conjugate of the expression to the left. The Bloch equations then reduce to
\begin{eqnarray} \label{Bloch_plus}
\frac{\partial M_+}{\partial t} &=& M_+\left(i\Delta\omega -\frac{1}{T_2}\right)+i\gamma h(\omega t)e^{-i\omega t}M_z, \\ \label{Bloch_minus}
\frac{\partial M_-}{\partial t} &=& M_-\left(-i\Delta\omega -\frac{1}{T_2}\right)-i\gamma h^*(\omega t)e^{i\omega t}M_z, \\
\frac{\partial M_z}{\partial t} &=& \frac{-M_z}{T_1}-\frac{i\gamma}{2}\left[h(\omega t)e^{-i\omega t}M_--h^*(\omega t)e^{i\omega t}M_+\right],
\end{eqnarray}
where $\Delta\omega\equiv\omega_0-\omega$. Since Eq.~\ref{Bloch_minus} is simply the complex conjugate of Eq.~\ref{Bloch_plus}, it is only necessary to solve for one of them. It can also be assumed that ${M_z>>M_+,M_-}$, since optical pumping in the $z$-direction will create a much larger magnetization than that created by the relatively weak strength of the RF field. Then the equation for $M_z$ has the solution $M_z=M_0e^{-t/T_1}$, in the case that the equilibrium magnetization is much smaller than $M_0$. This is the case after the pump beam is turned off. Defining $P_\perp=\frac{M_{x'}+iM_{y'}}{M_0}$,
\begin{eqnarray} \label{P_perp}
\frac{\partial P_\perp}{\partial t}=P_\perp\bigg(i\Delta\omega-\frac{1}{T_2}\bigg)+i\gamma h(\omega t)e^{-i\omega t}e^{-t/T_1}.
\end{eqnarray}

A magnetic wave with arbitrary polarization can be written as the sum of two oppositely rotating circularly polarized components, $B_L$ and $B_R$. A separate coordinate system can be defined for each magnetometer, with a common $x$-axis in the direction of the probe beams and the $z$-axis pointing in the direction of the local static field, so that $\hat{y}_1=-\hat{y}_2$ and $\hat{z}_1=-\hat{z}_2$. Then the RF magnetic field can be written in phasor notation as 
\begin{eqnarray} \label{B1}
\widetilde{\textbf{B}}_{1}=B_Le^{i\phi}\left(\hat{x}-i\hat{y}_1\right)+B_Re^{-i\phi}\left(\hat{x}+i\hat{y}_1\right)
\end{eqnarray}
for one of the two magnetometers, with $\hat{z}_1=\hat{z}$. For the other magnetometer,
\begin{eqnarray} \label{B2}
\widetilde{\textbf{B}}_{2}=B_Le^{i\phi}\left(\hat{x}+i\hat{y}_2\right)+B_Re^{-i\phi}\left(\hat{x}-i\hat{y}_2\right).
\end{eqnarray}
Here, $2\phi$ is the difference in phase between the right- and left-rotating magnetic fields. There can also be a common phase, but this is equivalent to a shift in time, and so it is not expressed explicitly. From the $x$- and $y$-components of these expressions for the magnetic field,
\begin{eqnarray} \label{h1}
h_1(\omega t) &=& B_Le^{i\omega t}e^{i\phi}+B_Re^{-i\omega t}e^{i\phi},\\ \label{h2}
h_2(\omega t) &=& B_Re^{i\omega t}e^{-i\phi}+B_Le^{-i\omega t}e^{-i\phi}.
\end{eqnarray}

Writing Eq.~\ref{P_perp} separately for the two magnetometers using Eqs.~\ref{h1}--\ref{h2} and using the secular approximation for $\omega$ close to $\omega_0$,
\begin{eqnarray}
\frac{\partial P_{\perp1}}{\partial t} &=& P_{\perp1}\bigg(i\Delta\omega-\frac{1}{T_2}\bigg)+i\gamma B_Le^{i\phi}e^{-t/T_1}, \\
\frac{\partial P_{\perp2}}{\partial t} &=& P_{\perp2}\bigg(i\Delta\omega-\frac{1}{T_2}\bigg)+i\gamma B_Re^{-i\phi}e^{-t/T_1}.
\end{eqnarray}
Defining $\Gamma \equiv \frac{1}{T_2}-\frac{1}{T_1}$, the solutions to these differential equations are
\begin{eqnarray}
P_{\perp1} &=& \frac{i\gamma B_Le^{i\phi}}{\Gamma-i\Delta\omega}(e^{-t/T_1}-e^{-t/T_2}e^{i\Delta\omega t})+P_{\perp1}^0e^{-t/T_2}e^{i\Delta\omega t},\\[10pt]
P_{\perp2} &=& \frac{i\gamma B_Re^{-i\phi}}{\Gamma-i\Delta\omega }(e^{-t/T_1}-e^{-t/T_2}e^{i\Delta\omega t})+P_{\perp2}^0e^{-t/T_2}e^{i\Delta\omega t},
\end{eqnarray}
where $P_{\perp1}^0$ and $P_{\perp2}^0$ are the initial values of the transverse polarization. Then if the time the data is acquired is much greater than $T_2$, the solutions are
\begin{eqnarray}
P_{\perp1} &\approx& \left(\frac{i\gamma e^{-t/T_1}}{\Gamma-i\Delta\omega}\right)B_Le^{i\phi},\\
P_{\perp2} &\approx& \left(\frac{i\gamma e^{-t/T_1}}{\Gamma-i\Delta\omega}\right)B_Re^{-i\phi}.
\end{eqnarray}
The magnetometer's signal, using phase-sensitive detection, is a direct measure of $P_\perp$.

From Eqs.~\ref{B1}-\ref{B2}, the $x$- and $y$-components of the RF magnetic field are given by 
\begin{eqnarray}
\widetilde{B}_{x} &=& B_Le^{i\phi}+B_Re^{-i\phi}, \\
\widetilde{B}_{y} &=& i(B_Re^{-i\phi}-B_Le^{i\phi}).
\end{eqnarray}
Since $P_{\perp1}\propto B_Le^{i\phi}$ and $P_{\perp2}\propto B_Re^{-i\phi}$, where the proportionality constants are in principle the same, the $x$- and $y$-components of the RF magnetic field can be found from the measurements taken by the two magnetometers using the relation 
\begin{eqnarray} \label{x_comp}
\widetilde{B}_{x} &\propto& P_{\perp1}+P_{\perp2}, \\ \label{y_comp}
\widetilde{B}_{y} &\propto& i(P_{\perp2}-P_{\perp1}).
\end{eqnarray}
To account for potential differences in the proportionality constants between the two sensors, the magnetometer's response can be calibrated with respect to a known reference signal. The absolute direction of the field can therefore be found by adding and subtracting the measurements from the two magnetometers with their tuning fields equal and opposite and probe beams along the same direction. Table~\ref{tab:mag_jones} shows the magnetic Jones vectors obtained from Eqs.~\ref{x_comp}-\ref{y_comp} for linear, right-circularly polarized, and left-circularly polarized magnetic fields. As shown in the table, the $x$- and $y$-components of linear magnetic fields will only have a real component, and the $x$- and $y$-components of rotating magnetic fields will be $90^\circ$ out of phase. For circularly polarized fields, the sensor detects the field in the co-rotating case but detects nothing in the counter-rotating case. Only half the amplitude of the field is detected by each sensor for linearly polarized fields.

    \begin{table}[h]
        \centering
        \setlength{\tabcolsep}{20pt}
        \renewcommand{\arraystretch}{1}
        \caption{Magnetic Jones vectors from the two signals are shown for five polarization states. $B$ is the ampltitude of the magnetic field. The quantities listed under the second and third columns are proportional to the measurements from the two magnetometers. Combining these quantities according to Eqs.~\ref{x_comp}-\ref{y_comp} gives the magnetic Jones vector.}
        \begin{tabular}{ccccc}
        \toprule
        Polarization State & Signal 1 & Signal 2 & Magnetic Jones Vector \\
        \midrule
        Linear along $x$ & $\frac{B}{2}$ & $\frac{B}{2}$ & $B\begin{bmatrix} 1 \\ 0 \\ \end{bmatrix}$ \\
        Linear along $y$ & $i\frac{B}{2}$ & $-i\frac{B}{2}$ & $B\begin{bmatrix} 0 \\ 1 \\ \end{bmatrix}$ \\
        Linear along $\phi$ & $\frac{B}{2}e^{i\phi}$ & $\frac{B}{2}e^{-i\phi}$ & $B\begin{bmatrix} \cos\phi \\ \sin\phi \\ \end{bmatrix}$ \\
        RCP & $0$ & $B$ & $B\begin{bmatrix} 1 \\ i \\ \end{bmatrix}$ \\
        LCP & $B$ & $0$ & $B\begin{bmatrix} 1 \\ -i \\ \end{bmatrix}$ \\
        \end{tabular}
        \label{tab:mag_jones}
    \end{table}

\maketitle

\section{Experiment}

\begin{figure}[h]
    \centering
    \includegraphics[trim=150 150 150 200, clip, width=\linewidth, valign=t]{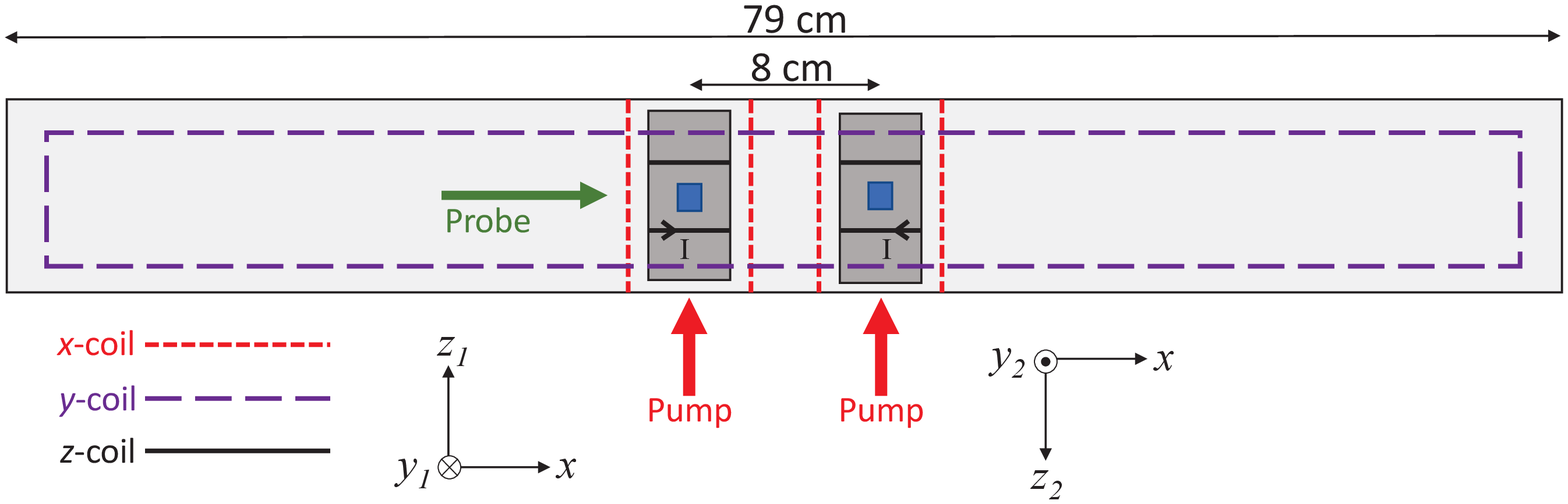}
    \caption{This schematic shows a top view of the sensors, with atomic cells shown in blue, and the field coils used in the experiment. The probe beams were in the $x$-direction, and pulsed pump beams were along $z=z_1$. RF fields were created using the $x$- and $y$-coils, which were wound around two plates separated by 2 cm. The equal and opposite currents shown in the schematic represent the equal and opposite tuning fields of the two magnetometers.}
    \label{fig:schematic}
\end{figure}

\begin{figure}[h]
        \centering
        \includegraphics[trim=30 50 100 50, clip, width=\linewidth, valign=t]{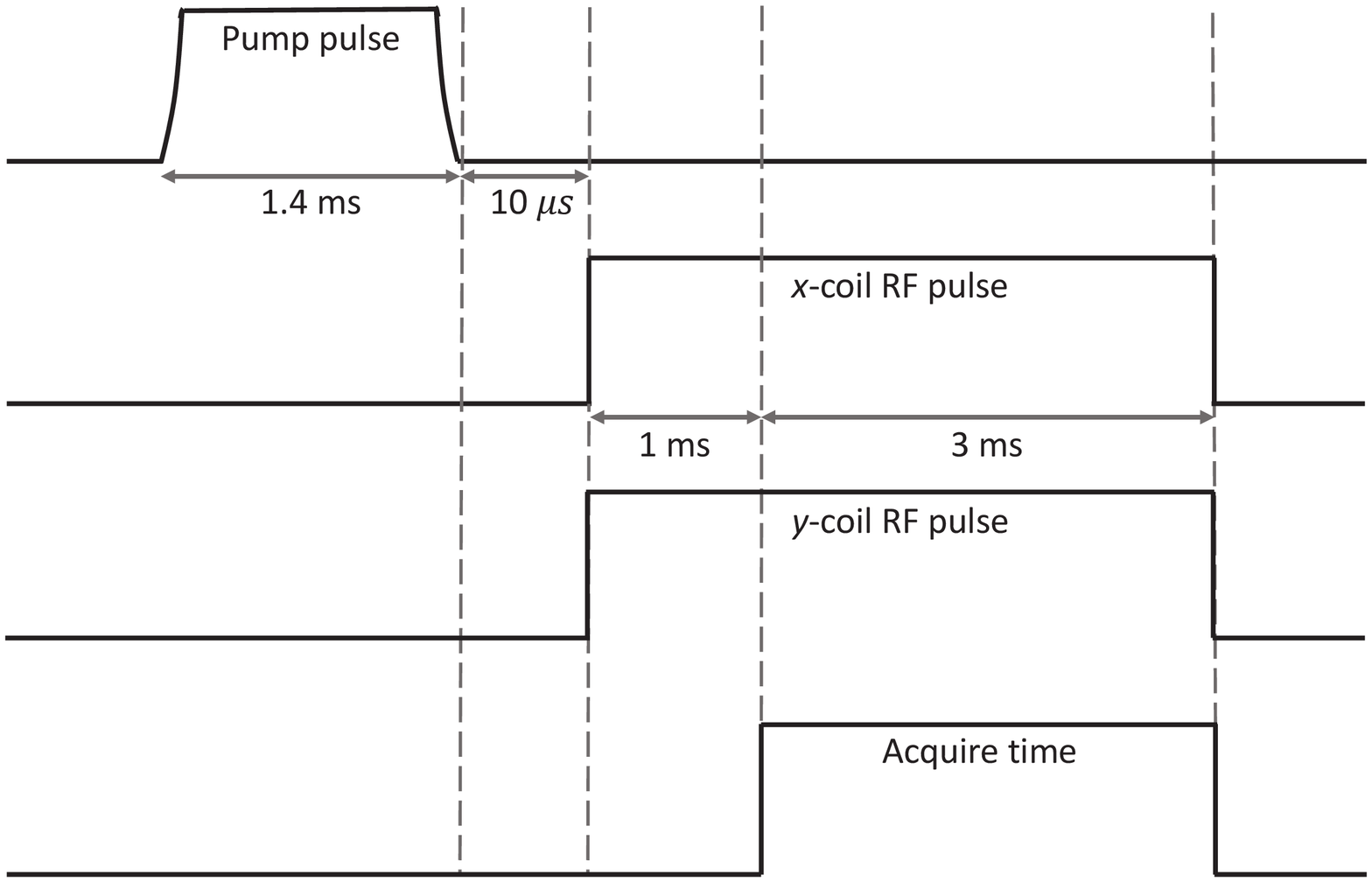}
        \caption{The RF fields were turned on 10 $\mu$s after the 1.4 ms pump pulse. The data was acquired for 3 ms, starting 1 ms after the RF fields were turned on.}
        \label{fig:sequence}
    \end{figure}

    A schematic of the experimental setup is shown in Fig.~\ref{fig:schematic}. Two multi-pass~\cite{yao2022,vachaspati2011,cooper2016} Twinleaf atomic magnetometers~\cite{Twinleaf,quiroz2022} were used, each with a glass cell containing rubidium-87. Neon with a number density of 0.8 amg was used as a buffer gas, and N$_2$ at 0.06 amg was used as a quenching gas. Each magnetometer had its own set of static field coils for canceling Earth's field and for creating gradient fields for improving the $B_z$ field homogeneity. The $B_z$ coils were used to produce the tuning fields, which were made to be equal in magnitude and opposite in direction for the two magnetometers. A circularly-polarized pump laser pointing in the $z$-direction was used in pulse mode for optical pumping, and a linearly polarized probe laser was used for the measurement. Each sensor had a probe laser that passed through the cell more than 30 times for signal amplification~\cite{quiroz2022}. A balanced polarimeter measured the rotation of the probe beam's polarization, and a phase-sensitive spectrometer from Tecmag~\cite{Tecmag} operating with a dwell time of 12 $\mu$s was used to collect the data. Fig.~\ref{fig:sequence} shows the experimental sequence.

The coordinate system was defined so that the $z$-axis points in the direction of one of the two tuning fields, and the $x$-axis points in the direction of the average of the probe beam pass directions. The orthogonality of the $x$- and $z$-axes was experimentally enforced within each sensor by using a static $x$-field to adjust the tuning field direction until zero DC magnetization was measured in the $x$-direction after the pump beam pulse. Nevertheless, misalignment between the axes of the two sensors could have resulted in errors in the field measurement. For instance, a field in the $x$-direction as defined by the probe beam direction for one of the sensors would result in a nominal $y$-field on the order of $B\varphi /2$, where $B$ is the amplitude of the applied field and $\varphi$ is the small angle between the two sensors' $x$-axes.

RF signals were created by running an alternating current through two calibrated coils. One of these coils, shown in red in Fig.~\ref{fig:schematic}, was designed to produce uniform fields in the $x$-direction. The other, shown in purple, produced uniform fields in the $y$-direction. The RF field coils used in the experiment were designed to be homogeneous over the two sensors spaced 8 cm apart, a distance large enough to prevent the separate tuning fields from significantly interfering with each other. The outputs of the spectrometer were used as the voltage sources for the two coils, and the difference in phase between the two components was varied to get different types of magnetic field polarization. Experiments were performed for linearly, circularly, and elliptically polarized magnetic waves. In addition, signals were made with linear and circular polarization at three different frequencies: on-resonance, off-resonance by 325 Hz, and off-resonance by 650 Hz.
    
    To fine-tune the field sizes, the counter-rotating case was created for one of the sensors by setting the x- and y-components 90$^{\circ}$ out of phase. To make the $x$- and $y$-components equal for the circularly polarized test signals, the $x$-component field was adjusted to minimize the counter-rotating signal. This method was then repeated for the other sensor to confirm that the signal was minimized in the counter-rotating case for both sensors.

\maketitle

\section{Results}
\begin{figure}[h]
        \centering
        \includegraphics[trim=0 160 0 170, clip, width=\linewidth, valign=t]{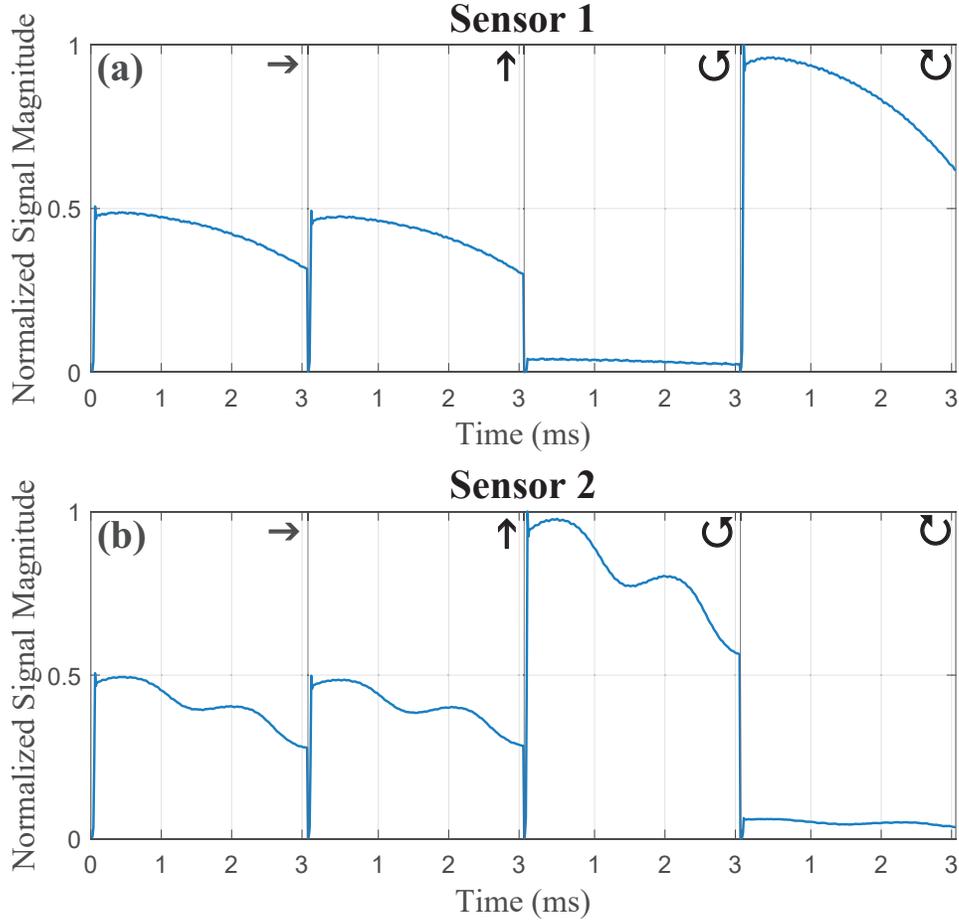}
        \caption{Co-rotating magnetic fields are preferentially detected by (a) sensor 1 and (b) sensor 2 while counter-rotating fields are suppressed. The symbols above the plot of each measurement show the polarization state of the resonant RF field. Slight misalignment of fields, by a few degrees, explains the residual signal seen by the nominally counter-rotating fields.}
        \label{fig:sensor_time}
    \end{figure}

    Fig.~\ref{fig:sensor_time} shows data from the first set of test signals made at the resonance frequency of 423.2 kHz, the NQR frequency of ammonium nitrate. Zero on the horizontal axis in Fig.~\ref{fig:sensor_time} corresponds to the start of the acquire time, as shown in Fig.~\ref{fig:sequence}. RF magnetic fields were created with directions along the $x$-axis, along the $y$-axis, rotating counterclockwise, and rotating clockwise, as indicated by the arrows. A magnetic field of $64\pm 5$ pT was used for both the $x$- and $y$-components of each test signal. The plots show the normalized magnitude of the complex signal from each measurement. Since each sensor was sensitive to fields co-rotating with the precession frequency, the plots for both sensors show a strong signal for the co-rotating field but a greatly suppressed signal for the counter-rotating field. The residual signal in the counter-rotating case can be explained by a slight misalignment of the $x$- and $y$-coils for creating the RF fields. If the angle between the direction of fields produced by the $y$-coil and the direction perpendicular to the $x$-coil field is $\theta$, then the ratio expected between the residual signal in the counter-rotating case and the signal in the co-rotating case is on the order of $\theta/2$ in the small angle approximation. The residual signal size of approximately {5\%} visible in Fig.~\ref{fig:sensor_time} suggests a misalignment between the coils of approximately $6^{\circ}$.

\begin{figure}[h]
        \centering
        \includegraphics[trim=0 160 0 170, clip, width=\linewidth, valign=t]{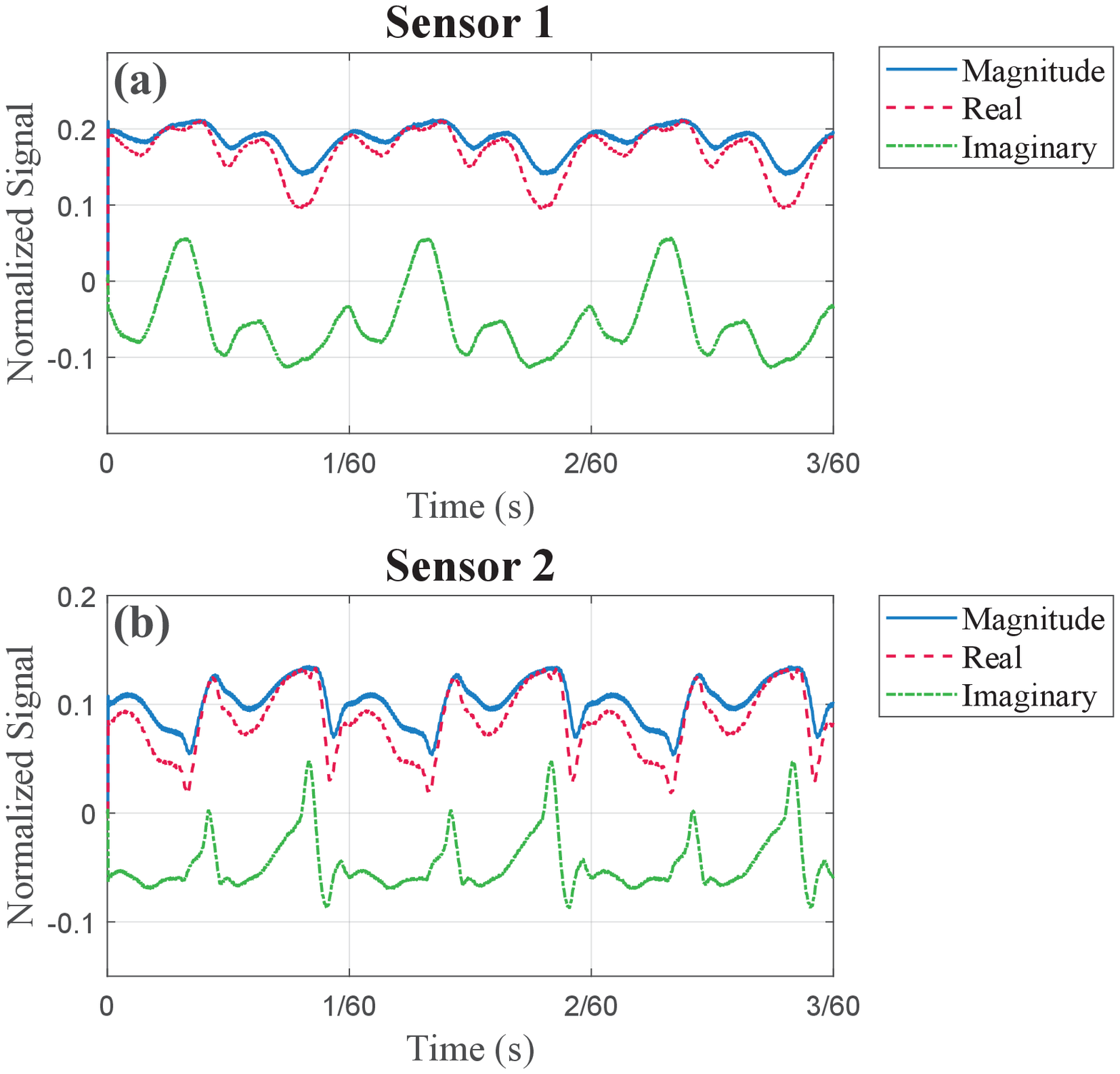}
        \caption{Steady-state data, taken with a longer acquire time, is shown for (a) sensor 1 and (b) sensor 2. Both sensors show a 60 Hz repeating pattern due to the power line. This 60 Hz pattern may explain the structure of the time domain signal shown in Fig.~\ref{fig:sensor_time}.}
        \label{fig:60Hz}
    \end{figure}

The structure, visible especially in the data from sensor 2, may be explained in part by the 60 Hz noise from the power line. To show the 60 Hz pattern of the signal's shape, the data plotted in Fig.~\ref{fig:60Hz} was taken with a longer acquire time, and the pump light was kept on at a low power throughout the measurement to achieve a steady state between the pumping and the decay. Three cycles of the 60 Hz pattern are visible in the plots for each sensor shown in Fig.~\ref{fig:60Hz}. To minimize the effect of the 60 Hz noise, the experimental sequence was triggered on the power line. Without triggering on the line, large variations were seen in the results for different repititions of the same experiment, but the signal became stable after it was synced to the power line.

    \begin{figure}[h]
        \centering
        \includegraphics[trim=0 170 0 170, clip, width=15cm, valign=t]{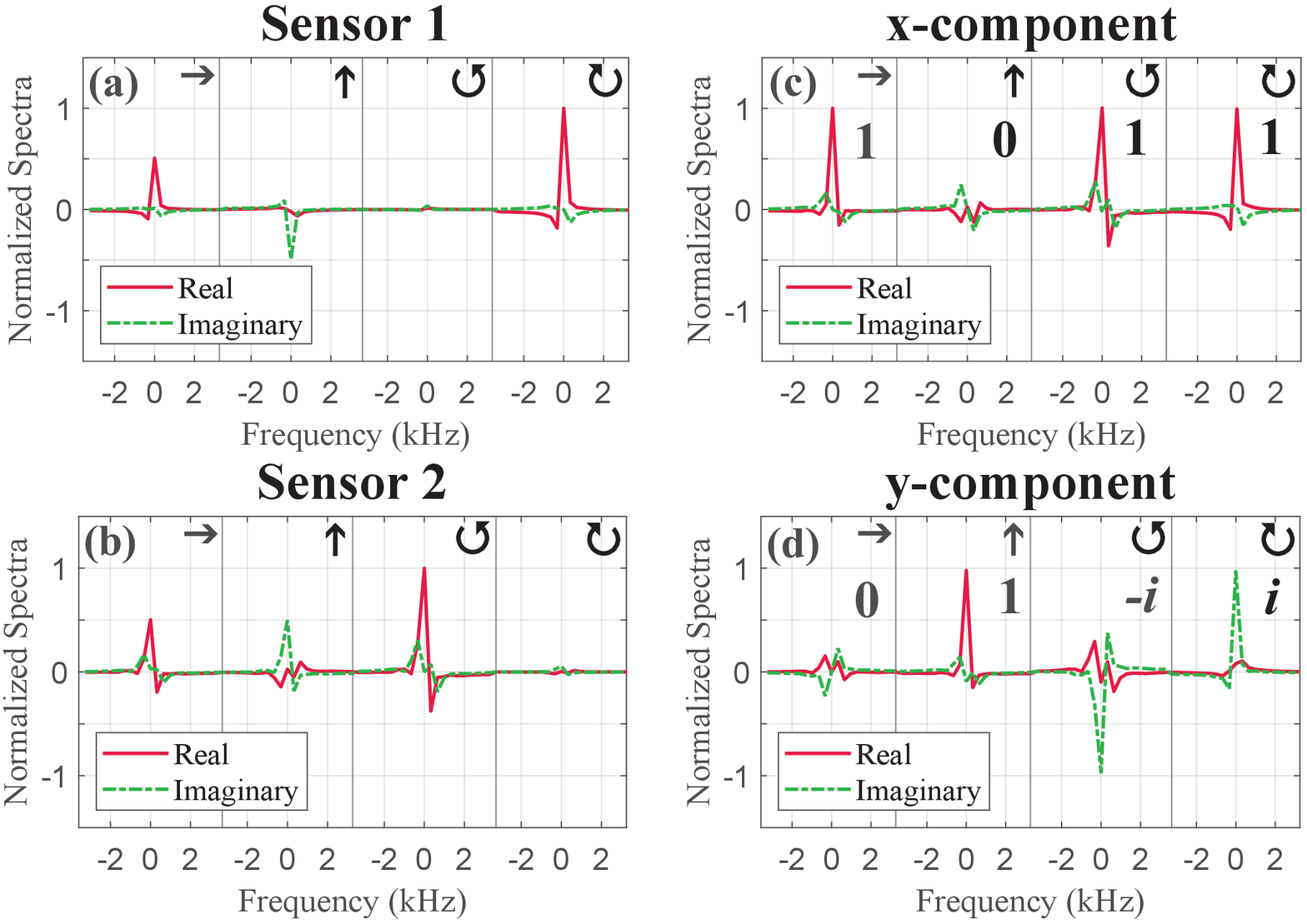}
        \caption{On the left, the Fourier transform of the time domain data displayed in Fig.~\ref{fig:sensor_time} is shown for (a) sensor 1 and (b) sensor 2. The plots on the right show the $x$- and $y$-components of the RF fields obtained using Eqs.~\ref{x_comp}-\ref{y_comp} from the frequency domain data shown on the left. The $x$-axis shows the difference in frequency from the Larmor frequency of 423.2 kHz. It was demonstrated that two sensors with tuning fields in opposite directions can be used to determine the absolute direction of the field without knowing the source. The on-resonance peaks in plots (c) and (d) give the magnetic Jones vector for each of the magnetic fields in the experiment.}
        \label{fig:freq_to_comp}
    \end{figure}

        A Fourier transform of the time domain data was taken, and the resulting frequency domain data were normalized to the $x$-component signal. The normalized frequency domain data are plotted in (a) and (b) of Fig.~\ref{fig:freq_to_comp}. In the frequency domain, it is again visible that the signals from the co-rotating fields have twice the amplitude of the linear signals. As expected for the two sensors with opposite tuning fields, the $y$-axis signals from the two sensors have opposite signs. The signal obtained from the $x$-axis field is real, while the $y$-axis signal is imaginary. This is due to the fact that the $y$-axis magnetic field results in precession of the atoms that is $90^{\circ}$ out of phase from the precession that results from a field along the $x$-direction.

        The direction of each test signal was determined using Eqs.~\ref{x_comp}-\ref{y_comp}. The result is shown in (c) and (d) of Fig.~\ref{fig:freq_to_comp}. The plots correctly show the $x$- and $y$-components expected for each magnetic field that was created. For the $x$- and $y$-axis fields, there is a real component in the direction expected. The small nonzero signal for the other component may be explained by 60 Hz noise, misalignment of the RF fields, or slight misalignment between the axes of the two sensors. However, these residual signals are close to zero on resonance. For the two rotating fields, the $x$-component is real while the $y$-component is imaginary, correctly representing the fact that the $x$- and $y$-components are $90^{\circ}$ out of phase. The $y$-component is either negative or positive depending on the sense of rotation. The on-resonance peaks in the component plots shown in (c) and (d) give the magnetic Jones vector for each test signal. Table~\ref{tab:my_label} shows the ideal magnetic Jones vector for each of the test signals represented in Fig.~\ref{fig:freq_to_comp} in the order they are presented in the plots. The constant $B$ represents the amplitude of the $x$- and $y$-components, and it can be complex depending on the origin of time. The Jones vectors obtained by adding and subtracting the data in the frequency domain, according to Eqs.~\ref{x_comp}-\ref{y_comp}, are in agreement with the directions expected for the magnetic fields based on the amplitude and phase chosen for the fields from the $x$- and $y$-coils.

    \begin{table}[h]
        \centering
        \setlength{\tabcolsep}{20pt}
        \renewcommand{\arraystretch}{1}
        \caption{The ideal magnetic Jones vectors are listed for the first set of test signals shown in Fig.~\ref{fig:freq_to_comp}. The quantity $B$ is the amplitude of the magnetic field.}
        \begin{tabular}{ccccc}
        \toprule
        Test Signal&1&2&3&4\\
        \midrule
        Jones Vector
        &$B\begin{bmatrix} 1 \\ 0 \\ \end{bmatrix}$
        &$B\begin{bmatrix} 0 \\ 1 \\ \end{bmatrix}$
        &$B\begin{bmatrix} 1 \\ -i \\ \end{bmatrix}$
        &$B\begin{bmatrix} 1 \\ i \\ \end{bmatrix}$
        \end{tabular}
        \label{tab:my_label}
    \end{table}

    \begin{figure}[h]
        \centering
        \includegraphics[trim=0 160 0 160, clip, width=\linewidth, valign=t]{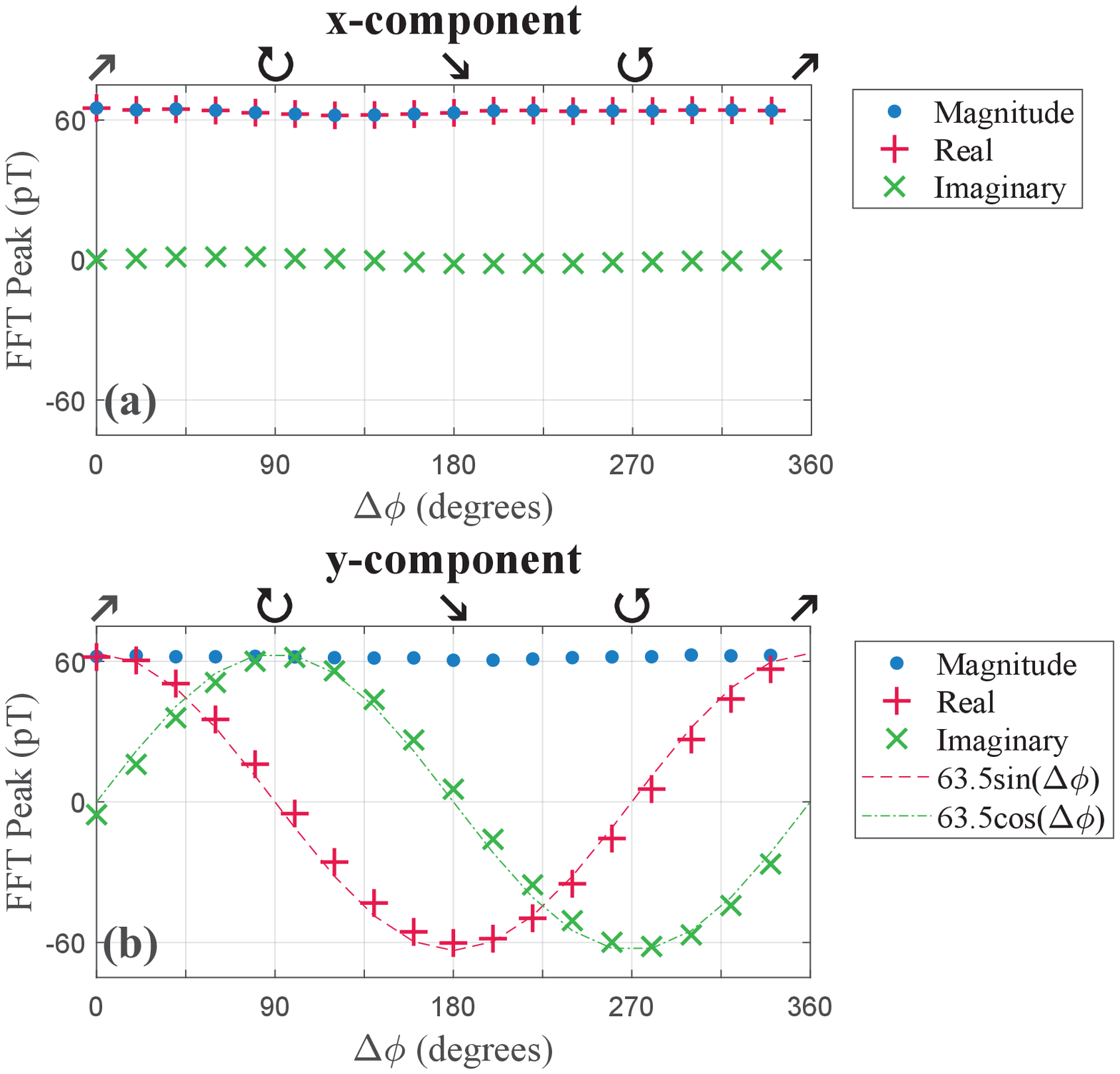}        
        \caption{As the polarization state is varied by incrementing the phase of the $y$-component, $\Delta\phi$, while leaving the phase of the $x$-component fixed, the corresponding Jones vector is traced out by combining signals from the two sensors, according to Eqs.~\ref{x_comp}-\ref{y_comp}. These plots demonstrate the ability to determine the degree of circular polarization using two sensors with their tuning fields in opposite directions.}
        \label{fig:20_deg}
    \end{figure}

After demonstrating the ability to use two magnetometers to determine the absolute direction of linearly and circularly polarized magnetic fields, another experiment was performed to demonstrate the ability to determine the degree of circular polarization. In this experiment, the test signals were made by incrementing the phase of the $y$-component through $360^{\circ}$ while keeping the $x$-component unchanged. This corresponds to a magnetic Jones vector of
\begin{eqnarray} \label{20_deg_Jones}
\vec{B}=B\begin{bmatrix} 1 \\ e^{i\Delta\phi} \end{bmatrix},
\end{eqnarray}
where $\Delta\phi$ is the difference in phase between the $x$- and $y$-components. Fig.~\ref{fig:20_deg} shows the FFT peak from each test signal, after combining the signals from the two sensors according to Eqs.~\ref{x_comp}-\ref{y_comp}, plotted against the phase difference between the $x$- and $y$-components. The points in the graphs  therefore represent the components of the magnetic Jones vector. As the phase of the $y$-component is changed, the real and imaginary parts of the $y$-component change sinusoidally, as expected from Eq.~\ref{20_deg_Jones}. The arrows above the plot indicate how the polarization of the magnetic wave moves from linear through elliptical to circular polarization as the phase difference between the components changes.

To demonstrate that this method of determining the absolute direction of RF magnetic fields works for off-resonance fields as well, provided they are within the range of frequencies detectable by the magnetometer as determined by the linewidth of the sensors, a third experiment was performed using off-resonance test signals. This is important for applications where the target frequency is either unknown or known only within a certain range. In some applications, the resonance frequency drifts in time. For example, in nuclear quadrupole resonance, the frequency shifts with temperature.

    \begin{figure}[h]
        \centering
        \includegraphics[trim=0 160 0 160, clip, width=\linewidth, valign=t]{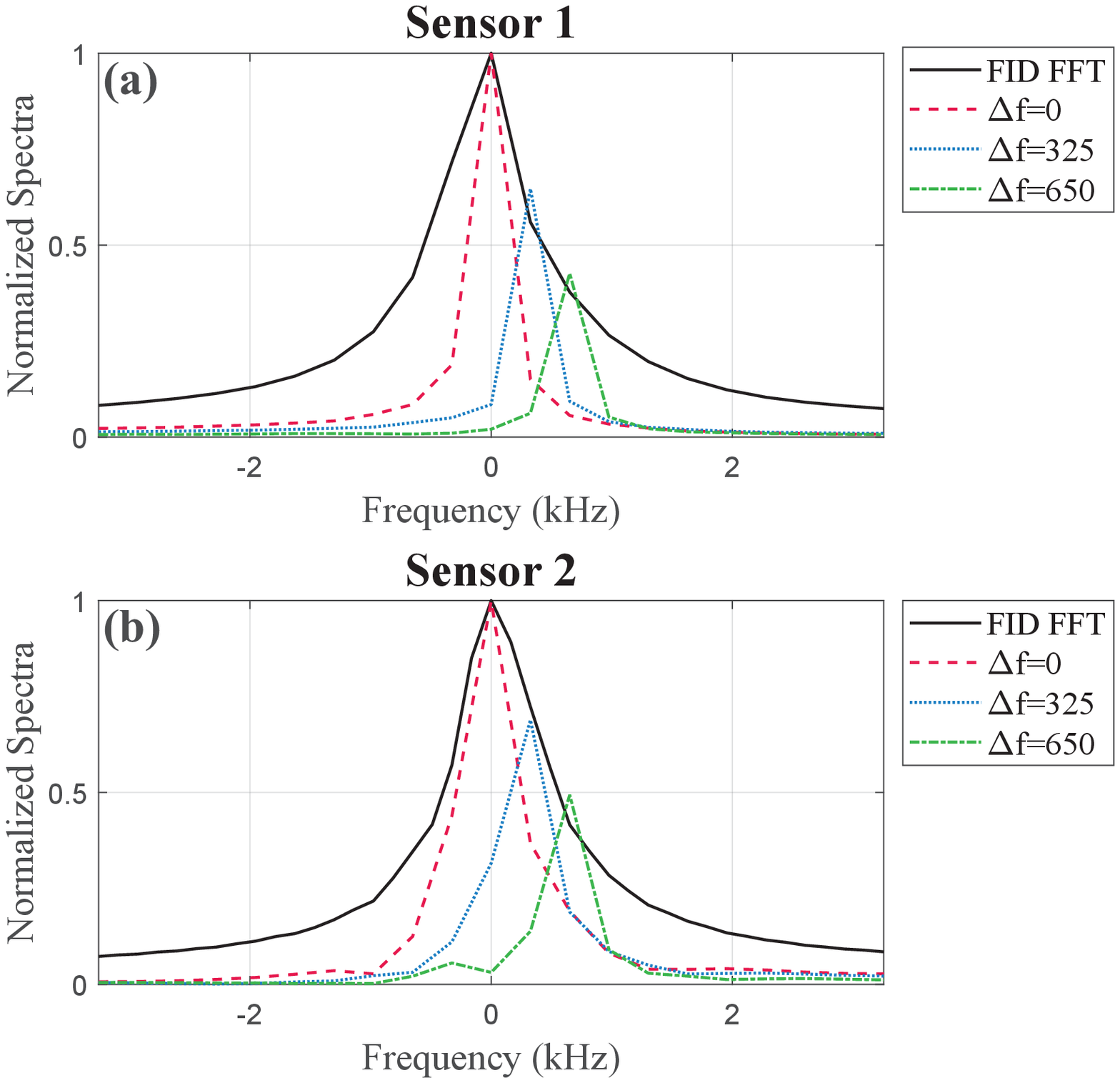}
        \caption{The frequency domain data from signals made along the $x$-axis at three different frequencies are plotted together with the Fourier transform of an FID signal. The legend shows the difference, $\Delta f$, between the frequency of the test signal and the resonance frequency of 423.2 kHz. As the signals are made to be more off-resonance, the amplitude decreases according to the spectral distribution of the sensor.}
        \label{fig:linewidth}
    \end{figure}

    \begin{figure}[h]
        \centering
        \includegraphics[trim=60 0 60 0, clip, width=\linewidth, valign=t]{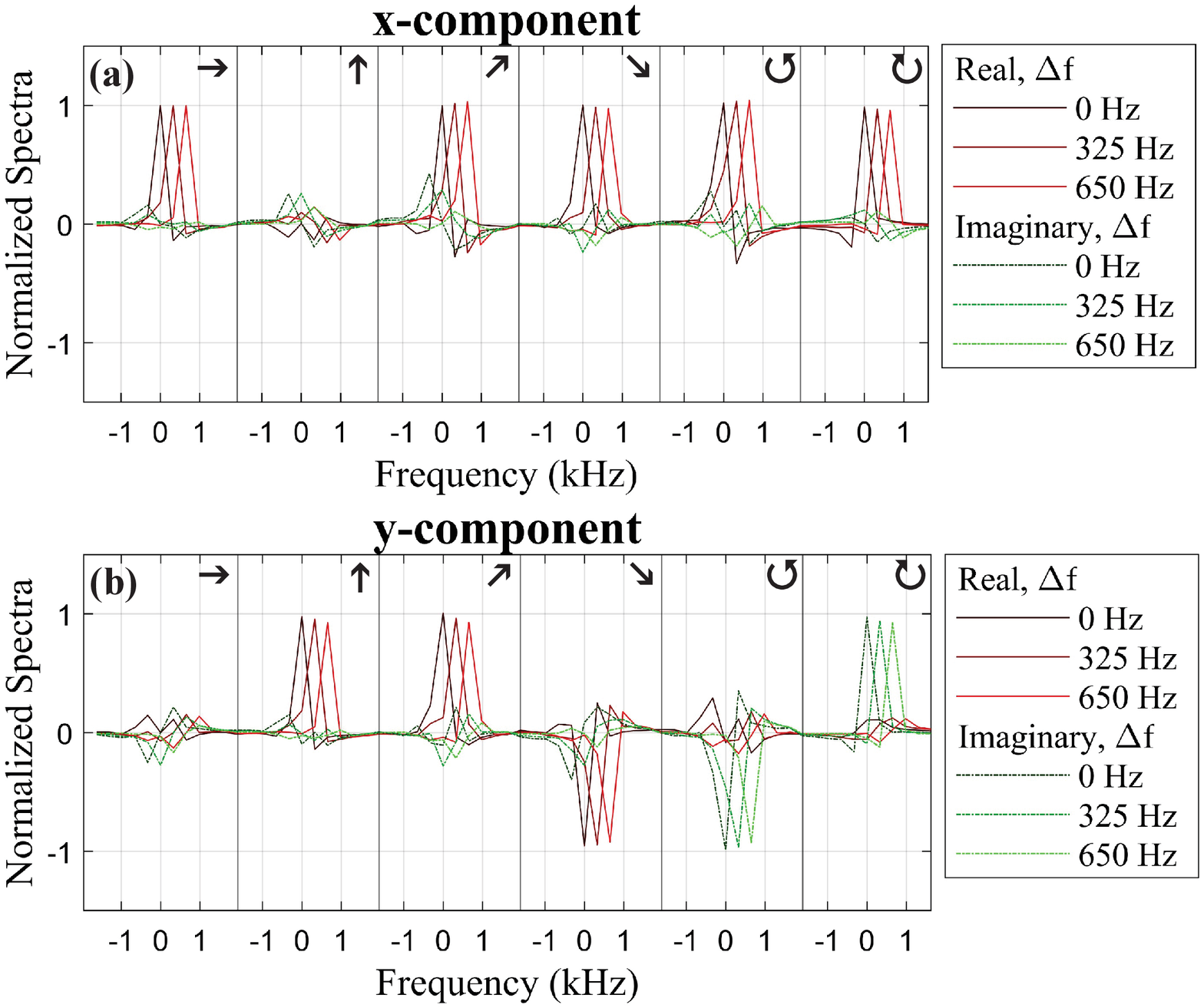}
        \caption{The $x$- and $y$-components of RF fields created at three different frequencies are plotted as a function of frequency. The correct $x$- and $y$-components were obtained even when the RF field was made to be off-resonance.}
        \label{fig:off_resonance}
    \end{figure}

The test signals in the third experiment were created at three different frequencies. For each direction of the magnetic field, the test signals were made to be on-resonance, off-resonance by 325 Hz, and off-resonance by 650 Hz. The normalized frequency domain data from the test signals created along the $x$-axis are plotted for both sensors in Fig.~\ref{fig:linewidth}, together with the Fourier transform of a free induction decay (FID). The amplitude of the signal decreases as the frequency is made to be more off-resonance, according to the spectral distribution of the sensor.

At each of the three frequencies, test signals were made along the $x$-axis, along the $y$-axis, along a line 45$^{\circ}$ up from the $x$-axis, along a line 45$^{\circ}$ down from the $x$-axis, rotating counterclockwise, and rotating clockwise. The Fourier transform of the data was taken, and the data at each frequency was normalized separately. Fig.~\ref{fig:off_resonance} shows plots of the $x$- and $y$-components found from the normalized frequency domain data. The components are plotted against the frequency, so the off-resonance signals are shifted to the right from the on-resonance signals. For all three frequencies, the $x$- and $y$-components obtained from the data agree with the result expected based on the amplitude and phase chosen for the magnetic fields from the $x$- and $y$-coils.

\maketitle

\section{Conclusion}

It was demonstrated that, due to a magnetometer's sensitivity to the polarization state of an RF magnetic field, the absolute direction and polarization of an RF field within a 2D plane could be determined using two magnetometers with their tuning fields equal and opposite. The described method was shown to work for off-resonance signals as well. Furthermore, the measurements could be expressed as a magnetic Jones vector for the RF field. This method of determining the direction and polarization of RF fields could be used to distinguish signal from noise, which would be particularly useful in an unshielded environment. Magnetic communications and radio direction finding could also benefit from the ability to determine the polarization state of an RF field. Uniquely determining the magnetic Jones vector of an RF field provides a new tool for characterizing these signals.

\section{Acknowledgments}

This research was funded in part by NSF (Award No.
1711118).  One of us (Motamedi) was supported by the Office of Student Scholarship, Creative Activities, and Research (OSCAR) at GMU through the Undergraduate Research Scholars Program (URSP).


%

\end{document}